# QUCoughScope: An Artificially Intelligent Mobile Application to Detect Asymptomatic COVID-19 Patients using Cough and Breathing Sounds

[1]Muhammad E. H. Chowdhury, [1]Nabil Ibtehaz, [1]Tawsifur Rahman, [2]Yosra Magdi Salih Mekki, [3]Yazan Qibalwey, [1]Sakib Mahmud, [1]Maymouna Ezeddin, [2]Susu Zughaier, [3]Sumaya Ali S A Al-Maadeed

[1]Electrical Engineering Department, College of Engineering, Qatar University, Doha 2713, Qatar
[2]College of Medicine, Qatar University, Doha 2713, Qatar
[3]Department of Computer Science and Engineering, College of Engineering, Qatar University, Doha 2713, Qatar

*Abstract*— In the break of COVID-19 pandemic, mass testing has become essential to reduce the spread of the virus. Several recent studies suggest that a significant number of COVID-19 patients display no physical symptoms whatsoever. Therefore, it is unlikely that these patients will undergo COVID-19 test, which increases their chances of unintentionally spreading the virus. Currently, the primary diagnostic tool to detect COVID-19 is reverse-transcription polymerase chain reaction (RT-PCR) test on collected respiratory specimens from the suspected case, which is physically invasive and must be performed by a trained clinician. This requires patients to travel to a laboratory facility to be tested, thereby potentially infecting others along the way. It is evident from recent researches that asymptomatic COVID-19 patients cough and breath in a different way than the healthy people. This motivated researcher to investigate cough and breath sound to distinguish COVID-19 patients from the patients with non-COVID lung infections and healthy population. Several research groups have created mobile and web-platform for crowdsourcing the symptoms, cough and breathing sounds from healthy, COVID-19 and Non-COVID patients. Some of these data repositories were made public. We have received such a repository from Cambridge University team under data-sharing agreement, where we have cough and breathing sound samples for 582 and 141 healthy and COVID-19 patients, respectively. 87 COVID-19 patients were asymptomatic, while rest of them have cough. We have developed an Android application to automatically screen COVID-19 from the comfort of people's homes. Test subjects can simply download a mobile application, enter their symptoms, record an audio clip of their cough and breath, and upload the data anonymously to our servers. Our backend server converts the audio clip to spectrogram and then apply our state-of-the-art machine learning model to classify between cough sounds produced by COVID-19 patients, as opposed to healthy subjects or those with other respiratory conditions. The result is then returned to the test user in the mobile application. The system can detect asymptomatic COVID-19 patients with a sensitivity more than 91%. Therefore, this system can be used by the patients' in their premises as a pre-screening method to aid COVID-19 diagnosis.

## I. INTRODUCTION

The novel coronavirus-2019 (COVID-19) disease is an acute respiratory syndrome that has already caused over 2.53 million causalities and infected more than 114 million people, as of March 01, 2021 [1]. The business, economic, and social dynamics of the whole world were affected. Governments have imposed flight restrictions, social distancing, and increasing awareness of hygiene. However, COVID-19 is still spreading at a very rapid rate. The common symptoms of coronavirus include fever, cough, shortness of breath, and pneumonia. Severe cases of the coronavirus diseases include result in acute respiratory distress syndrome (ARDS) or complete respiratory failure, which requires support from mechanical ventilation and an intensive-care unit (ICU). People with a compromised immune system, or elderly people are more likely to develop serious illnesses, including organs heart and kidney failures, particularly kidneys or and septic shocks [2].

The current diagnosis of COVID-19 is done by time-consuming, expensive, and in-accessible Reverse-Transcription Polymer Chain Reaction (RT-PCR) testing. This kit is not easily available in some regions due to lack of adequate supplies, medical professionals and healthcare facilities. Recently, Artificial Intelligence (AI) has been implemented in the health sector widely. There are so many applications in the field of speech and audio, AI has created its space. It is used for the screening and early detection of the different disease, which could help in saving a large number of lives. It is stated that breathing, speech, sneezing, and coughs can be used by machine learning models to diagnose different respiratory illness such as COVID-19 [3], [4], [5]. Thus, it is possible to identify whether a person is infected by the virus or not utilizing the respiratory signals like breathing and cough sounds.

In this work, we propose using cough and breathing sound for COVID-19 diagnosis, AI based model via smartphone application. This application can be used as a pre-screening tool to decrease the pressure on health centers, provide a faster and more reliable testing mechanism to reduce the spread of the virus. Our contribution can be summarized as follow:

- Conduct a literature review of related works to prove the potential applicability of the proposed solution.
- Point out the limitations of related works and how the proposed solution may overcome those problem.
- Trained, validated and tested state-of-the-art machine learning models and reported a pipeline as an innovative solution.
- Experimentally prove cough and breath sounds combination have latent features to distinguish COVID-19 patient from Non-COVID-19.
- An Android based smartphone application with backend server is created that allows the user to share

symptoms and cough and breath data for the COVID-19 diagnosis anonymously.

- To the best of our knowledge, "**QUCoughScope**" is the first solution which is not just a mobile application to collect crowd data rather we have implemented a deep-learning pipeline in the backend to immediately provide the screening outcome to the mobile phone application.

This article consists of six sections, in the introduction we explained the problem of the current COVID-19 testing approach and how it can be reduced with the help of our pre-screening tool. Section II highlights the related works while Section III introduces the methodology and Section IV explains the implementation details. Section V summarizes the system performance while Section VI concludes the article.

## II. RELATED WORK

In this section, some related work is presented briefly, to show that using smartphone based pre-screening tool, "QUCoughScope" is a potential COVID-19 detection tool.

Different body signals like respiration or heart signals were used by the research to automatically detect different lungs and heart diseases (for example for wheeze detection in asthma [6, 7, 8]). Human voice has been used to early detect several diseases like Parkinson's disease, coronary artery disease, traumatic brain injury and brain disorders. Parkinson's disease was linked to the softness of speech which can be resulting from lack of vocal muscle coordination [9, 10]. Different voice parameters like voice frequency, vocal tone, pitch, rhythm, rate, and volume can be correlated with coronary artery disease [11]. Invisible illnesses like post-traumatic stress disorder [12], traumatic brain injury and psychiatric conditions [13] can be linked with audio information. The use of human-generated audio can be used as a biomarker for the early detection of different disease and can be cheap solution for mass population or a screening and pre-screening tool. This is even become more useful and comfortable to the user if that is related to the daily activities and the data acquisition can be done non-invasively

Recent work has started exploring how respiratory sounds (e.g., coughs, breathing and voice) collected by devices from patients tested positive for COVID-19 in hospital differ from sounds from healthy people. Digital stethoscope data from lung auscultation is used as a diagnostic signal for COVID-19 [14], while 48 COVID-19 patients versus other pathological coughs collected with phones were used in a study to detect COVID-19 is presented using ensemble of CNN models [15]. In [16], speech recordings from COVID-19 hospital patients to analyzed automatically the health status of the patients. Our work contains an exploration of using human respiratory sounds as distinctive markers for COVID-19 using the crowdsourced data by Cambridge university group.

AI4COVID-19 is a mobile app which records 3 seconds of cough audio which is analysed automatically to provide an indication of COVID-19 status within 2 minutes proposed by Imran et al. [17] using cough sound and transfer learning. The pipeline is consisted of two stages, cough detection and collection, and COVID-19 diagnosis. starting with cough detection engine, user must record 3 seconds good quality cough sound, Mel spectrogram image of the wave is analyzed with fully connected CNN. After the cough is detected system pass to the COVID-19 diagnosis to decide the result. It consists of 3 AI approach, the Deep transfer learning Multi-class classifier (DTL-MC), the Classical machine learning Multi-class classifier (CML-MC) and deep transfer learning binary-class classifier. Some key limitations of the current AI4Covide-19 are: 1) limited training data, 2) limited data to generalize the model, 4) Cough features of COVID-19 may overlap with other diseases.

A medical dataset containing 328 cough sounds have been recorded from 150 patients of four different types: COVID-19, Asthma, Bronchitis and Healthy. A deep neural network (DNN) was shown to distinguish between COVID-19 and other coughs with an accuracy of 96.83% [18]. There appear to be unique patterns in COVID19 coughs that allow a pre-trained Resnet18 classifier to identify COVID-19 coughs with an AUC of 0.72. In this case cough samples were collected over the phone from 3621 individuals with confirmed COVID-19 [19]. COVID-19 coughs were classified with a higher AUC of 0.97 (sensitivity = 98.5% and specificity = 94.2%) by a Resnet50 architecture trained on coughs from 4256 subjects and evaluated on 1064 subjects that included both COVID-19 positive and COVID19 negative subjects [20].

Brown et al. [21] collected both cough and breathing sound, then investigated how such data can aid with COVID diagnosis. They provide a handcraft features for cough and breathing sounds such as duration, onset, tempo, period, root mean square (RMS) energy, spectral centroid, roll-off frequency, zero-crossing, Mel-frequency cepstrum (MFCC), delta MFCC and delta delta MFCC. Combined with deep transfer learning, VGGish, which is a convolution network designed to extract audio features automatically. They achieved a 0.80±0.7 accuracy with 2 class classification problem using the cough and breathing data. This dataset was shared to our team under data-sharing agreement, which was used to develop the machine learning pipeline and can be validated on Qatari data.

## III. METHODOLOGY

This section describes the dataset, data-preprocessing, experiments, evaluation parameters.

DATASET

Several public datasets are available such as Coswara [22], CoughVid [23] and Cambridge dataset [21]. However, the Cambridge dataset was not completely public, the team has made it available upon request. Among the accessible dataset Cambridge dataset was most reliable as that was acquired in a well-designed framework.

The Cambridge dataset was designed for developing a diagnostic tool for COVID-19 based on cough and breathing sounds [21]. The dataset was collected through an app (Android and Web (www.covid-19-sounds.org)) that asked volunteers for samples of their coughs and breathing as well as their medical history and symptoms. Age, gender, geographical location, current health status and preexisting medical conditions are also recorded. Audio recordings were sampled at 44.1 KHz and subjects were from different parts of the world. Cough and breathing sound samples were collected for 582 and 141 healthy and COVID-19 positive patients, respectively. Among them, 264 healthy and 54 COVID-19 patients have cough symptoms while 318 healthy and 87 COVID-19 patients have no symptoms (Table 1).

Table 1: Details of the Cambridge Dataset

| Health State | Breath | Cough | Forced cough | Total |
|---|---|---|---|---|
| Healthy | 582 | 264 | 318 | 582 |
| COVID-19 infected | 141 | 54 | 87 | 141 |
| Total | 723 | 318 | 405 | 723 |

## DATA PREPARATION

Since the dataset was collected using the web and android platform, the dataset was organized to two groups first: Cough and breath. Then each of these groups were sub-dived into symptomatic and asymptomatic groups. Each of the symptomatic and asymptomatic breathing and cough sounds for COVID-19 and healthy groups were visualized in time domain to see potential differences among them (Figure 1).

## EXPERIMENTS

In our proposed solution we combine the cough and breath sound to overcome the limitation of some related works. We have therefore investigated two different pipelines with three-different combinations: Cough only, breath only and cough and breathing in-combination. The Cambridge dataset was used to develop and validate the prototype system which can be validated in Qatar with the local data. Table 2 shows the experimental pipeline for this porotype development.

Table 2: Experimental pipelines for this study

| Pipelines | COVID | Healthy |
|---|---|---|
| Pipeline I (Symptomatic) | a. Cough<br>b. Breath<br>c. Cough + Breath | a. Cough<br>b. Breath<br>c. Cough + Breath |
| Pipeline II (Asymptomatic) | a. Cough<br>b. Breath<br>c. Cough + Breath | a. Cough<br>b. Breath<br>c. Cough + Breath |

## WAVE TO SPECTROGRAM

A spectrogram is a visual representation of an audio signal that shows the evolution of the frequency spectrum over time. A spectrogram is usually generated by performing a Fast Fourier Transform (FFT) on a collection of overlapping windows extracted from the original signal. The process of dividing the signal in short term sequences of fixed size and applying FFT on those independently is called Short-time Fourier transform (STFT). The spectrogram is basically the squared magnitude of the STFT of the signal s(t) for a window width, w. These are the parameters used for STFT: n_fft = 2048, hop_length = 512, win_length = n_fft, window='hann'.

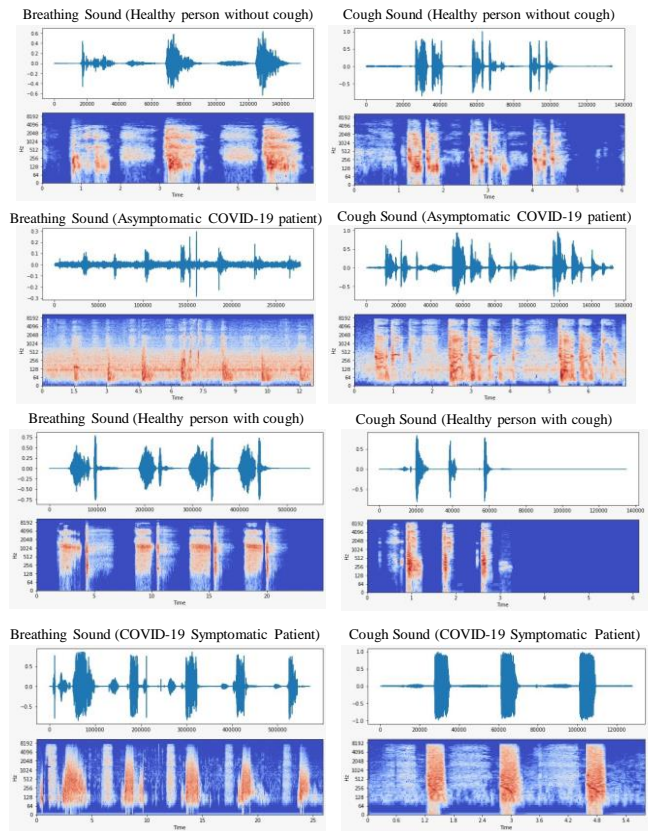

Figure 1: Breathing and cough sounds for asymptomatic and symptomatic healthy and COVID-19 patients.

## DATA AUGMENTATION

Table 1 shows that number of signals for both asymptomatic and symptomatic users are not equal and therefore the dataset is imbalance [24-27]. Therefore, to make the dataset balance for training the deep convolutional neural networks (CNNs), the training dataset was made balance using data augmentation technique.

Table 3: Number of mages per class and per fold used for different pipelines

| | Pipelines | # of Samples | Training Samples | Validation Samples | Testing Samples |
|---|---|---|---|---|---|
| Healthy | Asymptomatic | 318 | 229×11 =2519 | 25 | 64 |
| | Symptomatic | 264 | 190×13 =2470 | 21 | 53 |
| COVID-19 | Asymptomatic & symptomatic | 141 | 102×25 =2550 | 11 | 28 |
| | Symptomatic | 54 | 39×62 =2418 | 4 | 11 |

## EVALUATION METRICS

The performance of the COVID-19 detection was assessed using five evaluation metrics: Accuracy, Precision, Sensitivity, F1-score, and Specificity.

$$Accuracy = \frac{TP + TN}{TP + TN + FP + FN}$$

where $accuracy$ is the ratio of the correctly classified cases.

*TP, TN, FP, FN* represent the true positive, true negative, false positive, and false negative, respectively.

$$Precision = \frac{TP}{TP + FP}$$

where *precision* is the rate of correctly classified positive class CXR samples among all the samples classified as positive samples.

$$Sensitivity = \frac{TP}{TP + FN}$$

where *sensitivity* is the rate of correctly predicted positive samples in the positive class samples,

$$F1 = 2\frac{Precision \times Sensitivity}{Precision + Sensitivity}$$

where $F1$ is the harmonic average of precision and sensitivity.

$$Specificity = \frac{TN}{TN + FP} \quad (6)$$

where *specificity* is the ratio of accurately predicted negative class samples to all negative class samples.

PyTorch library with Python 3.7 was used to train and evaluate the deep CNN networks, with an 8-GB NVIDIA GeForce GTX 1080 GPU card. Adam optimizer was used, with initial learning rate, $\alpha = 10^{-4}$, an adaptive learning rate which decreases the learning parameter by a factor of 5 if validation loss did not improve for 3 consecutive epochs, early stopping criterion of 8 epochs, momentum updates, $\beta_1 = 0.9$ and $\beta_2 = 0.999$, and mini-batch size of 4 images with 40 backpropagation epochs.

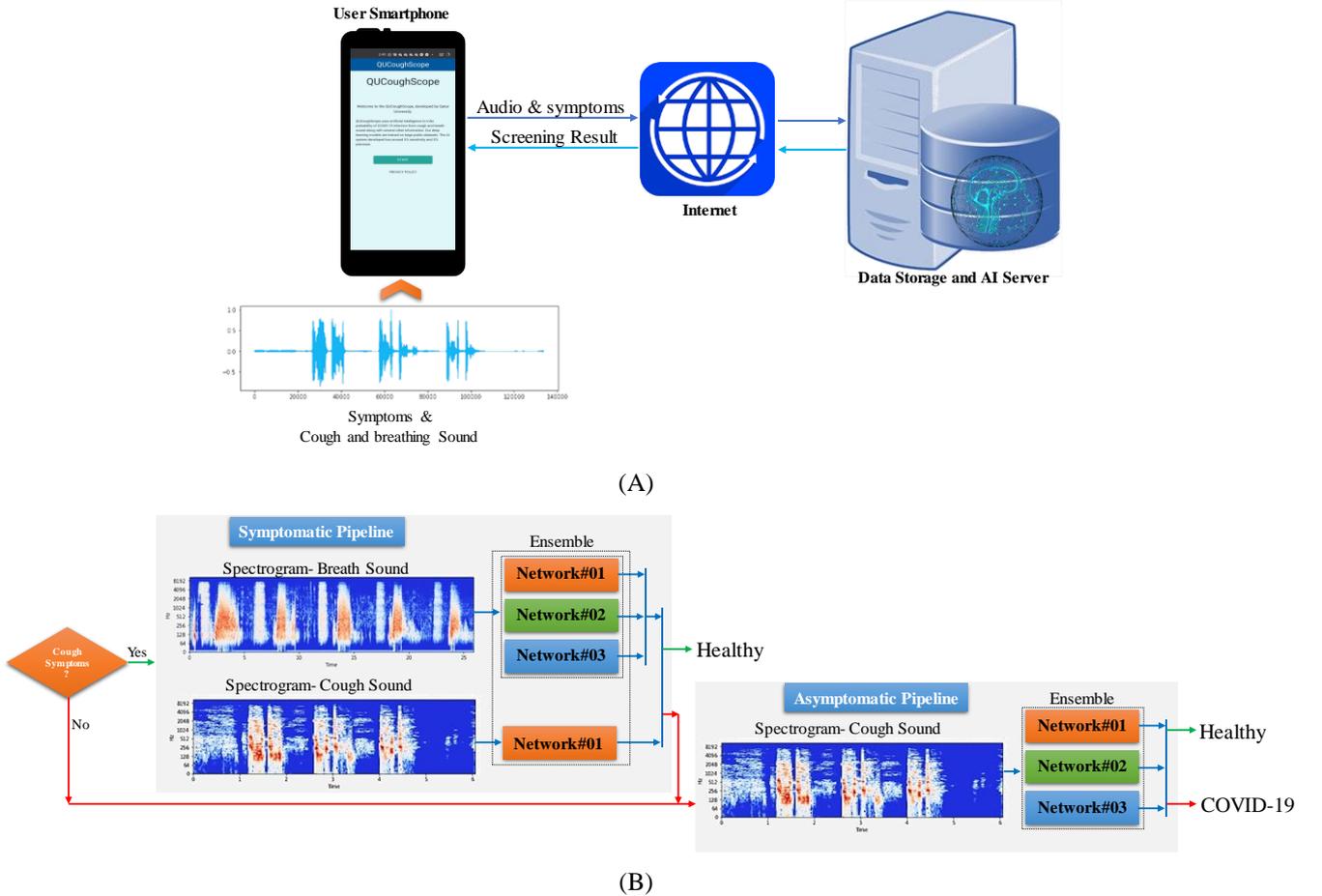

Figure 2: Illustration of generic framework for the QUCoughScope application (A) and the AI-based decision making pipeline (B).

## IV. IMPLEMENTATION

In the prototype system, the user will fill in some demographic data, as well as a list of confirmed symptoms. Next, once the app collects the cough and breathing sound data from the user, and it will be transferred to our server which perform signal processing and machine learning classification to determine whether your cough and breathing sounds like coronavirus patients or not (Figure 2A). Our app then notifies the users about their status. The mobile-recorded audio signal and symptoms once received by the sever machine, it performs STFT operation to convert raw audio signals into spectrogram images without any pre-processing.

Our pipeline is divide into two parts: symptomatic users (who has cough) and asymptomatic user (who doesn't have cough). Once the spectrogram is generated, our AI enabled server will check whether the user has cough or not based on that two separate pipelines it will carry out. If the user has entered that he/she has cough symptom, the symptomatic pipeline is activated. It was observed that for differentiating COVID and healthy symptomatic patients breathing sounds play important role than cough sounds. Three state-of-the-art deep learning models' output from breathing data will be ensemble with best performing network output from cough data to identify if the patient is healthy or normal. If the network identifies that the user is healthy user, it will report back the result to the user in our mobile application. However, if the network identifies the user is a potential COVID-19 patient, the decision will come after second stage of evaluation through our asymptomatic pipeline, which is very sensitive to COVID-19 detection (Figure 2B).

In the asymptomatic pipeline, the spectrogram image representing the cough sound will be fed to three state-of-the-art networks and their results will be ensemble to take decision of healthy and COVID-19 patients and the result will be reported back to mobile phone of the users. All these will be happened within couple of seconds.

*Architecture*

There are three state-of-the-art novel deep CNN models were deployed for this system and their results were ensemble to boost the overall performance further. The cascaded pipeline designed to increase the overall sensitivity of COVID-19 symptomatic patients is another novel framework of this system. This significantly reduces miss-classification and provides state-of-the-art performance.

We deployed our deep learning model pipelines as server applications. The server has been developed in Flask. Consequently, an android app has been developed using Flutter to acquire symptoms and audio data and the results are reported back to the mobile application. The mobile application "QUCoughScope" developed for this project can be accessed from this link: https://nibtehaz.github.io/qu-cough-scope/.

## V. PERFORMANCE

The following section illustrates the individual class performance and overall performance of the QUCoughScope. It is evident from the Table 4 that the novel framework developed by the Qatari team can stratify COVID-19 and healthy asymptomatic users with the sensitivity of 91.49% and 97.8%, respectively. The overall sensitivity of the system is 95.86%, whereas the sensitivity reported on this dataset by the Cambridge group was only 69% [21] - a marginal performance difference on the same dataset with our novel machine learning models and pipeline. In case of symptomatic users, our system can stratify COVID-19 and healthy user with the sensitivity of 81.48% and 95.08% respectively and overall sensitivity of 92.77% in comparison to 72% sensitivity of Cambridge group [21].

Most important contribution of this work is that every other group developed a mobile application that can only collect data [21,22,23, 28]. So far none of these groups have released an app which can actually acquire data and then give you a result within seconds at the comfort of your home. Our app is fully functional and can provides immediate diagnosis.

Table 4: Illustration of overall performance of the QUCoughScope system in detecting asymptomatic and symptomatic users.

**Cough Sound**

| Networks | COVID-19 (Asymptomatic and Symptomatic) | | | | | Healthy (Asymptomatic) | | | | | Overall | | | | |
|---|---|---|---|---|---|---|---|---|---|---|---|---|---|---|---|
| | Accuracy | Precision | Sensitivity | F1-score | Specificity | Accuracy | Precision | Sensitivity | F1-score1 | Specificity | Accuracy | Precision | Sensitivity | F1-score | Specificity |
| Network#01 | 93.90% | 99.13% | 80.85% | 89.06% | 99.69% | 93.90% | 92.15% | 99.69% | 95.77% | 80.85% | 93.90% | 94.29% | 93.90% | 93.71% | 86.64% |
| Network#02 | 94.34% | 94.57% | 86.52% | 90.37% | 97.80% | 94.34% | 94.24% | 97.80% | 95.99% | 86.52% | 94.34% | 94.34% | 94.33% | 94.26% | 89.99% |
| Network#03 | 95.20% | 92.86% | 92.20% | 92.53% | 96.86% | 95.42% | 96.55% | 96.86% | 96.70% | 92.20% | 95.42% | 95.42% | 95.43% | 95.42% | 93.63% |
| Ensemble | 95.86% | 94.85% | 91.49% | 93.14% | 97.80% | 95.86% | 96.28% | 97.80% | 97.04% | 91.49% | 95.86% | 95.84% | 95.86% | 95.84% | 93.43% |

**Breathing Sound**

| Networks | COVID-19 (Asymptomatic and Symptomatic) | | | | | Healthy (Asymptomatic) | | | | | Overall | | | | |
|---|---|---|---|---|---|---|---|---|---|---|---|---|---|---|---|
| | Accuracy | Precision | Sensitivity | F1-score | Specificity | Accuracy | Precision | Sensitivity | F1-score1 | Specificity | Accuracy | Precision | Sensitivity | F1-score | Specificity |
| Network#01 | 70.15% | 52.22% | 33.33% | 40.69% | 86.48% | 70.15% | 74.53% | 86.48% | 80.06% | 33.33% | 70.15% | 67.68% | 70.15% | 67.97% | 49.66% |
| Network#02 | 70.37% | 54.39% | 21.99% | 31.32% | 91.82% | 70.37% | 72.64% | 91.82% | 81.11% | 21.99% | 70.37% | 67.03% | 70.37% | 65.82% | 43.44% |
| Network#03 | 65.58% | 44.22% | 46.10% | 45.14% | 74.21% | 65.58% | 75.64% | 74.21% | 74.92% | 46.10% | 65.58% | 65.99% | 65.57% | 65.77% | 54.74% |

**Cough Sound**

| Networks | COVID-19 (Symptomatic) | | | | | Healthy (Symptomatic) | | | | | Overall | | | | |
|---|---|---|---|---|---|---|---|---|---|---|---|---|---|---|---|
| | Accuracy | Precision | Sensitivity | F1-score | Specificity | Accuracy | Precision | Sensitivity | F1-score | Specificity | Accuracy | Precision | Sensitivity | F1-score | Specificity |
| Network#01 | 81.76% | 47.73% | 77.78% | 59.16% | 82.58% | 81.76% | 94.78% | 82.58% | 88.26% | 77.78% | 81.76% | 86.79% | 81.76% | 83.32% | 78.60% |
| Network#02 | 77.99% | 42.00% | 77.78% | 54.55% | 78.03% | 77.99% | 94.50% | 78.03% | 85.48% | 77.78% | 77.99% | 85.58% | 77.99% | 80.23% | 77.82% |
| Network#03 | 78.62% | 41.03% | 59.26% | 48.49% | 82.58% | 78.62% | 90.83% | 82.58% | 86.51% | 59.26% | 78.62% | 82.37% | 78.62% | 80.05% | 63.22% |

**Breathing Sound**

| Networks | COVID-19 (Symptomatic) | | | | | Healthy (Symptomatic) | | | | | Overall | | | | |
|---|---|---|---|---|---|---|---|---|---|---|---|---|---|---|---|
| | Accuracy | Precision | Sensitivity | F1-score | Specificity | Accuracy | Precision | Sensitivity | F1-score | Specificity | Accuracy | Precision | Sensitivity | F1-score | Specificity |
| Network#01 | 92.45% | 75.00% | 83.33% | 78.95% | 94.32% | 92.45% | 96.51% | 94.32% | 95.40% | 83.33% | 92.45% | 92.86% | 92.45% | 92.61% | 85.20% |
| Network#02 | 90.88% | 68.66% | 85.19% | 76.04% | 92.05% | 90.88% | 96.81% | 92.05% | 94.37% | 85.19% | 90.88% | 92.03% | 90.89% | 91.26% | 86.35% |
| Network#03 | 89.62% | 64.79% | 85.19% | 73.60% | 90.53% | 89.62% | 96.76% | 90.53% | 93.54% | 85.19% | 89.62% | 91.33% | 89.62% | 90.15% | 86.10% |
| Ensemble* | 92.77% | 77.19% | 81.48% | 79.28% | 95.08% | 92.77% | 96.17% | 95.08% | 95.62% | 81.48% | 92.77% | 92.95% | 92.77% | 92.84% | 83.79% |

*Best network of cough-sound is ensemble with 3 best performing networks of breathing sound

## VI. CONCLUSION

One day, Qatar's citizens and residents will be able to breathe easier knowing that they can test safely at home for COVID-19 and other respiratory diseases. This AI-enabled application can help to reduce the spread of the COVID-19 virus and can also help beyond COVID19 detection. Respiratory syncytial virus (RSV) is a major viral infection of the respiratory system/ lungs among young children under the age of 2 years old. RSV infection could cause critically severe pneumonia that requires hospitalization. RSV affects breathing due to congestion and causes dry cough when mild infection; severe cough and wheezing in severe infection and critical illness children have difficulty in breathing. RSV infection is very common among young children, therefore an App that can detect RSV and distinguish it from asthma cough or wheezing sounds would be a highly beneficial medical application that helps medical doctors' decisions as management is very different between RSV and asthma or common cold or other bacterial infections like pertussis that also causes whooping cough sounds.